# RaceFixer - An Automated Data Race Fixer

Sanjay Malakar* | Tameem Bin Haider | Rifat Shahriar

Bangladesh University of Engineering and Technology, ,

**Correspondence**

*Corresponding author name, This is sample corresponding address. Email: iamsanjaymalakar@gmail.com

**Present Address**

This is sample for present address text this is sample for present address text

**Abstract**

Fixing software bugs has always been an essential and time-consuming process in software development. Fixing concurrency bugs has become especially critical in the multicore era. However, fixing concurrency bugs is challenging due to non-deterministic failures and tricky parallel reasoning. Beyond correctly fixing the original problem in the software, a good patch should also avoid introducing new bugs, degrading performance unnecessarily, or damaging software readability. Existing tools cannot automate the whole fixing process and provide good-quality patches. We present *RaceFixer*, a tool that automates the process of fixing one common type of concurrency bug: single-variable atomicity violations. RaceFixer starts from the bug reports of an existing bug-detection tool *ThreadSanitizer*. It augments these with static analysis to construct a suitable patch for each bug report. It tries to combine the patches of multiple bugs for better performance and code readability. Finally, we test RaceFixer on benchmarks from *TheadSanitizer*.

## 1 | INTRODUCTION

With the introduction of multi-core and other parallel architectures, there is an increased need for efficient and effective handling of software executing in such architectures. An essential aspect in this context is understanding the bugs that occur due to parallel and concurrent software execution. Concurrency bugs in multi-threaded programs have already caused real-world disasters and are a growing threat to software reliability in the multi-core era. Fixing bugs has always been a time-consuming and challenging part of software development, and concurrency bugs have become critical in our multi-core era. Concurrency bugs bring unique challenges, such as understanding data races, atomicity violations. A previous study of open-source software finds that it takes 73 days on average to fix a concurrency bug correctly. A study of operating-system patches shows that among common bug types, concurrency bugs are the most difficult to fix correctly.

Concurrency bugs can be categorized into several disjoint classes, such as data race, deadlock, order violation, etc. A data race is when two threads concurrently access a shared memory location, and at least one of the accesses is a write. Such races can lead to hard-to-reproduce bugs that are time-consuming to debug and fix. However, fixing concurrency bugs can be challenging because of non-deterministic failures and tricky parallel reasoning. Furthermore, apart from fixing the bugs, the patch must not introduce any new bugs and should not overly degrade the performance. A general technique to fix concurrency bugs is to insert new locks (known as gate locks) statically or dynamically to serialize all threads' executions. The inserted gate locks prevent two or more threads from executing concurrently eliminating the original incorrect thread interleaving. However, introducing gate locks may introduce performance bugs and various deadlocks.

Fixing the bugs can be divided into two parts- detecting the bugs and then fixing them. Detection of feasible races relies on detection of apparent data races. The three basic types of detection techniques are static, on-the-fly, and postmortem. On-the-fly and postmortem techniques refer to as dynamic. All three methods have their advantages and drawbacks.

Static techniques analyze the program information to report data races from source codes without any execution, while dynamic techniques locate data races from execution information of the program. Static detectors are sound, but imprecise since they report too many false positives. Dynamic detectors are precise or imprecise, but unsound since they cannot guarantee to locate the existence of at least one data race in a given execution of the program if there exists any. Dynamic detectors employ trace based post-mortem methods or on-the-fly methods, which report data races occurred in an execution of a programs. Post-mortem methods analyze the traced information or re-execute the program after an execution. On-the-fly methods are based on three different analysis methods: lockset analysis, happens-before analysis, and hybrid analysis.

Lockset analysis reports data races of monitored program by checking violations of a locking discipline, and happens-before analysis reports data races between current access and maintained previous accesses by comparing their happens-before relation based on the usage of a logical time stamp, such as vector clocks. The lockset analysis is simple and can be implemented with low overhead. However, lockset analysis may lead to many false positives, because it ignores synchronization primitives which are non-common lock such as signal/wait, fork/join, and barriers. The happens-before analysis is precise, since it does not report false positives and can be applied to all synchronization primitives. However, it is quite difficult to be efficiently implemented due to the performance overheads. The hybrid method tries to reduce the main drawback of pure lockset analysis and to get more improved performance than pure happens-before analysis.

Several dynamic data race detection techniques are performed in automatic tools with their significant advantages. These techniques used in detectors have some limitations, because they analyze only the dynamic execution of a program with a single input. Most dynamic detectors try to cover the limitations by considering the ordering of synchronization operations, such as fork-join, locks, signal-waits, and barriers, obtained in an actual execution of the program, but they still provide limited advantages (e.g. supporting particular synchronization primitives, improving the efficiency or the preciseness of execution overhead, etc).

We propose a tool RaceFixer for C programs to fix concurrency bugs. We mainly focus on fixing the bugs. As for detecting them, we are taking the help of a third-party tool ThreadSanitizer which is a part of the LLVM Clang project. RaceFixer uses static analysis and code transformation to insert locks and fix concurrency bugs detected by ThreadSanitizer.

## 2 | BACKGROUND

### 2.1 | Concurrency Bugs

Concurrency bugs can be classified into seven disjoint classes as follows[1]:

- **Data** race occurs when at least two threads access the same data and at least one of them write the data. It occurs when concurrent threads perform conflicting accesses by trying to update the same memory location or shared variable.

- **Deadlock** is "a condition in a system where a process cannot proceed because it needs to obtain a resource held by another process but it itself is holding a resource that the other process needs". More generally, it occurs when two or more threads attempts to access shared resources held by other threads, and none of the threads can give them up. During deadlock, all involved threads are in a waiting state.

- **Livelock** happens when a thread is waiting for a resource that will never become available while the CPU is busy releasing and acquiring the shared resource. It is similar to deadlock except that the state of the process involved in the livelock constantly changes and is frequently executing without making progress.

- **Starvation** is "a condition in which a process indefinitely delayed because other processes are always given preference". Starvation typically occurs when high priority threads are monopolising the CPU resources. During starvation, at least one of the involved threads remains in the ready queue.

- **Suspension-based locking or Blocking suspension** occurs when a calling thread waits for an unacceptably long time in a queue to acquire a lock for accessing a shared resource.

- **Order violation** is defined as the violation of the desired order between at least two memory accesses. It occurs when the expected order of interleavings does not appear. If a program fails to enforce the programmer's intended order of execution then an order violation bug could happen.

- ***Atomicity violation*** refers to the situation when the execution of two code blocks (sequences of statements protected by lock, transaction) in one thread is concurrently overlapping with the execution of one or more code blocks of other threads such a way that the result is inconsistent with any execution where the blocks of the first thread are executed without being overlapping with any other code block.

## 2.2 | LLVM

The LLVM Project is a collection of modular and reusable compiler and toolchain technologies. Used to develop a front end for any programming language and a back end for any instruction set architecture. LVM is written in C++ and is designed for compile-time, link-time, run-time, and "idle-time" optimization. It is being used in production by a wide variety of commercial and open source projects as well as being widely used in academic research.

### 2.2.1 | LLVM Core

The LLVM Core libraries provide a modern source-and target-independent optimizer, along with code generation support for many popular CPUs. These libraries are built around a well specified code representation known as the LLVM intermediate representation ("LLVM IR").

### 2.2.2 | Clang

Clang is an "LLVM native" C/C++/Objective-C compiler, which aims to deliver amazingly fast compiles, extremely useful error and warning messages and to provide a platform for building great source level tools. The Clang Static Analyzer and clang-tidy are tools that automatically find bugs in code, and are great examples of the sort of tools that can be built using the Clang frontend as a library to parse C/C++ code. There are various sanitizers available in Clang.

- AddressSanitizer (detects addressability issues)

- LeakSanitizer (detects memory leaks)

- ThreadSanitizer (detects data races and deadlocks) for C++ and Go

- MemorySanitizer (detects use of uninitialized memory)

- HWASAN, or Hardware-assisted AddressSanitizer, a newer variant of AddressSanitizer that consumes much less memory

- UBSan, or UndefinedBehaviorSanitizer

**ThreadSanitizer** (aka **TSan**) is a data race detector for C/C++.
TSan-LLVM detectable bugs :

- Normal data races

- Races on C++ object vptr

- Use after free races

- Races on mutexes

- Races on file descriptors

- Races on file descriptors

- Races on pthread_barrier_t

- Destruction of a locked mutex

- Leaked Threads

- And more...

## 3 | RELATED WORKS

The general approach in fixing atomicity violation is to introduce locks to protect the region of code that violates the atomicity assumption. However, several issues arise with locks: (1) can the approach identify atomic regions without being limited by issues such as the number of variables accessed, the number of instructions included, and the type of constructs, (2) can the approach select the appropriate locks and reuse locks if possible, (3) can the approach ensure that there are no deadlocks or data races introduced by the new locking mechanism, (4) does the solution to atomicity violation degrade performance, etc.

The current approaches to automatically fix atomicity violations face some of these issues. AFix[2] uses the result of other atomicity violation detection tools and introduce locks that create mutually exclusive regions. However, it applies to single variables, does not protect against deadlocks occurring in the fixed code, and introduces performance overhead. Grail[3] improves upon this by considering the context rather than only the buggy statements. Still, the fixes are not optimal and have performance problems. There is CFix[4], next version of AFix, with improved performance and deadlock avoidance. Understanding and Generating High Quality Patches for Concurrency Bugs[5] conducts an in-depth study of manual patches for 77 real-world concurrency bugs, which provides both assessments for existing techniques and actionable suggestions for future research. Guided by this study, a new tool HFix is designed. Adaptively Generating High Quality Fixes for Atomicity Violations[6] proposed alpha-Fixer to adaptively fix atomicity violations. Experimental results on 15 previously used atomicity violations show that: alpha-Fixer correctly fixed all 15 atomicity violations without introducing deadlocks. However, GLA and Grail both introduced 5 deadlocks. HFix only fixed 2 atomicity violations and introduced 4 deadlocks.

Automatically Fixing C Buffer Overflows Using Program Transformations[7] and Program Transformations to Fix C Integers[8] are both works on fixing bugs on C.

Although may works have been done in concurrency bug fixing but most of them are done as an optimization in an intermediate language and not in the C source language. Only alpha-Fixer[6] has done works regarding this. RaceFixer is also another one that provides fixes on the source level.

## 4 | PROPOSED RACEFIXER

*ThreadSanitizer of LLVM* (Dynamic Race Detection with LLVM Compiler Compile-time instrumentation for ThreadSanitizer) can be used to detect concurrency bugs. *TSan-LLVM*, a dynamic race detector that uses compile-time instrumentation, is based on a widely available LLVM compiler.

1. *TSan* can provide a list of concurrency bugs in a source code. After processing this list, *RaceFixer* can be provided with the locations to place the mutexes.

2. For each location, *RaceFixer* first defines the new mutex and then places the mutex around that location. There are several possible scenarios for this step as this depends on the type of statement on that location. Depending on the type of statement the mutex is placed accordingly.

3. Afterwards Step 1 is repeated on the changed source code(after adding the mutexes) as there might be more bugs found. This process is repeated until no concurrency bugs are found.

One thing to note here is that as *TSan* is a dynamic race detector, different bugs might be found in different runs. So a certain number of runs is required to build the report.

## 5 | METHODOLOGY

### 5.1 | Race Detection

There are a number of approaches to data race detection. The three basic types of detection techniques are: static, on the-fly and postmortem. On-the-fly and postmortem techniques are often referred to as dynamic. Static data race detectors analyze the source code of a program (e.g.[9]). When the code is large and complex, the performance of static detection degrades. Dynamic data race detectors analyze the trace of a particular program execution. On-the-fly race detectors process the program's events in parallel with the execution[10,11]. The postmortem technique consists in writing such events into a temporary file and then analyzing this file after the actual program execution[12]. Most dynamic data race detection tools are based on one of the following algorithms: happens-before, lockset or both (the hybrid type). ThreadSanitizer is a dynamic detector of data races. ThreadSanitizer uses a new algorithm; it has several modes of operation, ranging from the most conservative mode (which has few false positives but also misses real races) to a very aggressive one (which has more false positives but detects the largest number of real races). To the best of our knowledge ThreadSanitizer has the most detailed output and it is the only dynamic race detector with hybrid and pure happens-before modes. It has introduced the dynamic annotations, a sort of API for a race detector. Using the dynamic annotations together with the most aggressive mode of ThreadSanitizer enables us to find the largest number of real races while keeping zero noise level (no false positives or benign races are reported)[13].

So, we decided to use ThreadSanitizer as our bug detection tool.

## 5.2 | Parsing ThreadSanitizer Report

ThreadSanitizer reports a lot of information about the detected bugs.We wrote a parser to extract only the location and variable names where bugs are detected. An example ThreadSanitizer output might be like this :

```
WARNING: ThreadSanitizer: data race (pid=9376)
  Write of size 4 at 0x000000f29380 by thread T1:
      #0 Thread1 /home/sanjay/llvm/source/sample_codes/race.c:5:10 (a.out+0x4c7767)
  Previous write of size 4 at 0x000000f29380 by main thread:
      #0 main /home/sanjay/llvm/source/sample_codes/race.c:12:10 (a.out+0x4c77ae)
  Location is global 'Global' of size 4 at 0x000000f29380 (a.out+0x000000f29380)
  Thread T1 (tid=9378, running) created by main thread at:
      #0 pthread_create /home/sanjay/llvm/source/projects/compiler-rt/lib/tsan/rtl/
         ↪ tsan_interceptors.cc:967:3 (a.out+0x4614f1)
      #1 main /home/sanjay/llvm/source/sample_codes/race.c:11:3 (a.out+0x4c77a4)
SUMMARY: ThreadSanitizer: data race /home/sanjay/llvm/source/sample_codes/race.c:5:10 in Thread1
==================
ThreadSanitizer: reported 1 warnings
```

*Parser* generates a summary of the report ThreadSanitizer provides in 'VariableName Line Column' format. The above TSan log will be parsed to this :

```
Global 5 10 12 10
```

This means the data race is in a variable named *Global* and the two locations where data race happens is line 5 column 10 and line 12 column 10 of the source file.

## 5.3 | Source Transformation

For updating the source code we used clang's libTooling to source-to-source transformation. Clang AST is designed to be immutable. We can't change it after parsing. So state of the art source to source transformation in Clang requires the following steps:

1. Parse C++ source code to AST

2. Apply text replacements to original source code, generate new source code

3. Parse new source code to create new AST

## 5.4 | Race in Normal Statement

We checked every statement where a variable is referenced.

```
StatementMatcher varRef =
    stmt(hasDescendant(declRefExpr()),
        anyOf(hasParent(compoundStmt()), hasParent(ifStmt()), hasParent(whileStmt())))
    .bind("varRef");
```

If any statement contains a variable which ThreadSanitizer reported as potential race, we applied mutex lock and unlock before and after that statement.

### 5.4.1 | Race in If Condition

If the data race is in the condition of if statement three possible scenario can happen :

1. Race in if condition having else

```
if( RACE ){
....
}else{
....
}
```

For this case, RaceFixer puts a mutex lock before the if condition and releases the lock inside if conditional statement and else statement. The fixes produced by RaceFixer is underline below.

```
// lock added by RaceFixer
lock(mutex);
if( RACE ){
    // unlock added by RaceFixer
    unlock(mutex);
....
}else{
    // unlock added by RaceFixer
    unlock(mutex);
....
}
```

2. Race in if condition without having else

```
if( RACE ){
....
}
```

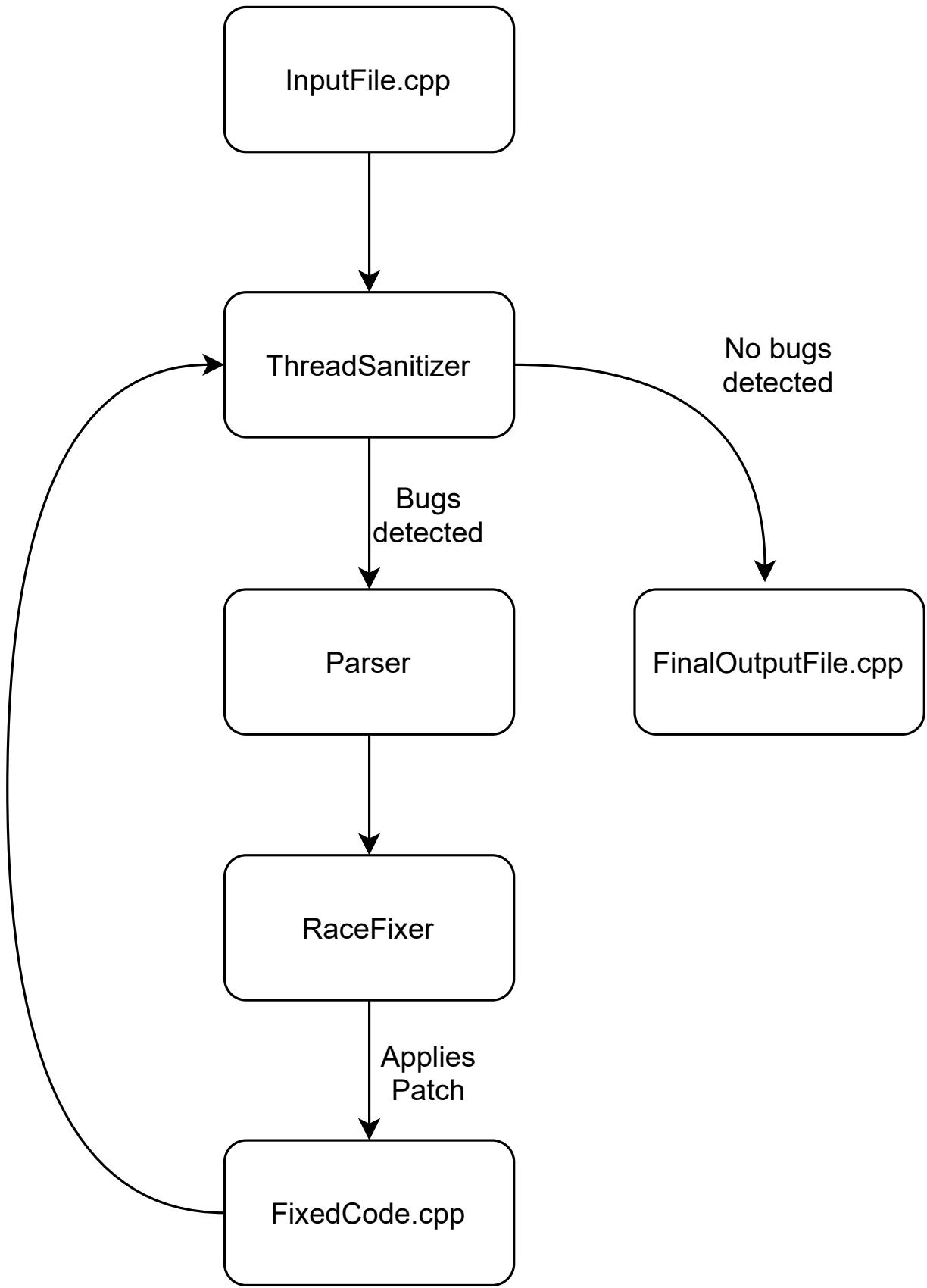

**FIGURE 1** Flow chart of RaceFixer

For this case, RaceFixer adds a else statement and puts a mutex lock before the if condition and releases the lock inside if conditional statement and else statement. The fixes produced by RaceFixer is underline below.

```
// lock added by RaceFixer
lock(mutex);
if( RACE ){
    // unlock added by RaceFixer
    unlock(mutex);
....
}else{
    // unlock added by RaceFixer
    unlock(mutex);
....
}
```

3. Race in else if condition

```
if( ... ){
....
}else if( RACE ) {
....
}
```

For this case, RaceFixer isolates the if condition alone. Then it splits the else if as a separate if statement. For the if statement generated from else if RaceFixer puts a mutex lock before the if condition and releases the lock inside if conditional statement and else statement. The fixes produced by RaceFixer is underline below.

```
// isolated if statement
if( ... ) {
    ....
}
// new if statement generated from else if
// lock added by RaceFixer
lock(mutex);
if( RACE ){
    // unlock added by RaceFixer
    unlock(mutex);
....
}else{
    // unlock added by RaceFixer
    unlock(mutex);
....
}
```

## 5.4.2 | Race in While Condition

If the race is in the condition of a while statement :

```
while ( RACE ){
....
}
```

RaceFixer fixes this by these steps :

- Lock before while

- Unlock in the start of while body

- Lock again in the end of while body

- Unlock after the while loop

```
// lock added by RaceFixer
lock(mutex);
while ( RACE ){
    // unlock added by RaceFixer
    unlock(mutex);
    ....
    // lock added by RaceFixer
    lock(mutex);
}
// unlock added by RaceFixer
unlock(mutex);
```

## 5.5 | Limitations

The main limitation of our work will be the bugs to be fixed i.e. *TSan*. The report from this tool lets us know about the locations of probable concurrency bugs. It also tells us about the acquired locks on that stage. But the only useful details would be the row and column number of bug in the source code. The rest of the details do not reference to source code as such they are not useful for our tool. Also there may possibly be false positives or false negatives on the report generated by *TSan*.

## 6 | CONCLUSION

We have described RaceFixer, a framework for automatically fixing a common type of concurrency bugs. We have implemented the system and shown RaceFixer to be effective at generating patches for atomicity-violation bugs detected by an TSan in several small benchmarks and a a real-world application. RaceFixer conducts thorough static analysis to reach a good balance among correctness, performance and code readability in its automatically generated patches. In the future we plan to extend RaceFixer to work with more general synchronization primitives and more types of concurrency bug detectors. Also from the types of bugs detectable by Tsan, there are many more available that may be used for various other tools and finally a more general bug fixing tool may be created.